\documentclass[letterpaper,12pt]{article}
\pdfoutput=1 
\usepackage{amsmath, latexsym, amsfonts, amssymb}
\usepackage{geometry, epsf}
\usepackage{graphicx}
\usepackage{epstopdf}
\newcommand  \gtsim  {\lower.5ex\hbox{$\; \buildrel > \over \sim \;$}} 
\newcommand  \ltsim  {\lower.5ex\hbox{$\; \buildrel < \over \sim \;$}}

\begin{document}
\vskip30pt
\bigskip
\centerline{\LARGE \bf Observing the Evolution of the Universe}
\centerline{\bf Cover page for a white paper in support of fine angular scale CMB probes.\footnote{This white paper was assembled by Lyman Page with input from many of the co-signers. It is part of the efforts of NASA'a Primordial Polarization Program Definition Team (PPPDT), Shaul Hanany chair, and of a NASA award to Steve Meyer and colleagues entitled ``A study for a CMB Probe of Inflation" (07-ASMCS07-0012).} }
\medskip
Co-signers in support of the science: James Aguirre, Alexandre Amblard,
Amjad Ashoorioon,
Carlo Baccigalupi,
Amedeo Balbi,
James Bartlett,
Nicola Bartolo,
Dominic Benford,
Mark	 Birkinshaw,
Jamie Bock,
Dick	Bond,
Julian Borrill,
Franois Bouchet,
Michael Bridges,
Emory Bunn,
Erminia Calabrese,
Christopher Cantalupo,
Ana Caramete,
Carmelita	Carbone,
Suchetana Chatterjee,
Sarah Church,
David Chuss,
Carlo Contaldi,
Asantha Cooray,
Sudeep Das,
Francesco De Bernardis,
Paolo De Bernardis,
Gianfranco De Zotti,
Jacques Delabrouille,
F.-Xavier	DŽsert,
Mark Devlin,
Clive	 Dickinson,
Simon Dicker,
Matt	Dobbs,
Scott	Dodelson,
Olivier Dore,
Jessie Dotson,
Joanna Dunkley,
Maria Cristina Falvella,
Dale	Fixsen,
Pablo Fosalba,
Joseph Fowler,
Evalyn Gates,
Walter Gear,
Sunil Golwala,
Krzysztof	Gorski,
Alessandro Gruppuso,
Josh	Gundersen,
Mark	 Halpern,
Shaul Hanany,
Masashi Hazumi,
Carlos Hernandez-Monteagudo,
Mark	Hertzberg,
Gary	Hinshaw,
Christopher Hirata,
Eric	Hivon,
Warren Holmes,
William Holzapfel,
Wayne Hu,
Johannes	Hubmayr,
Kevin Huffenberger,
Kent	Irwin,
Mark	 Jackson,
Andrew Jaffe,
Bradley Johnson,
William Jones,
Manoj Kaplinghat,
Brian Keating,
Reijo Keskitalo,
Justin Khoury,
Will	Kinney,
Theodore	Kisner,
Lloyd Knox,
Alan	Kogut,
Eiichiro Komatsu,
Arthur Kosowsky,
John	Kovac,
Lawrence	Krauss,
Hannu Kurki-Suonio,
Susana Landau,
Charles Lawrence,
Samuel Leach,
Adrian Lee,
Erik	Leitch,
Rodrigo Leonardi,
Julien Lesgourgues,
Andrew Liddle,
Eugene Lim,
Michele Limon,
Marilena	Loverde,
Philip Lubin,
Antonio Magalhaes,
Davide Maino,
Tobias Marriage,
Victoria Martin,
Sabino Matarrese,
John	 Mather,
Harsh Mathur,
Tomotake	Matsumura,
Pieter Meerburg,
Alessandro Melchiorri,
Stephan Meyer,
Amber Miller,
Michael Milligan,
Kavilan Moodley,
Michael Neimack,
Hogan Nguyen,
Ian	O'Dwyer,
Angiola Orlando,
Luca	 Pagano,
Lyman Page,
Bruce Partridge,
Timothy Pearson,
Hiranya Peiris,
Francesco Piacentini,
Lucio Piccirillo,
Elena Pierpaoli,
Davide Pietrobon,
Giampaolo Pisano,
Levon Pogosian,
Dmitri Pogosyan,
Nicolas Ponthieu,
Lucia Popa,
Clement Pryke,
Christoph	Raeth,
Subharthi	Ray,
Christian	Reichardt,
Sara	Ricciardi,
Paul	Richards,
Graca Rocha,
Lawrence	Rudnick,
John	 Ruhl,
Benjamin	Rusholme,
Claudia Scoccola,
Douglas Scott,
Carolyn Sealfon,
Neelima Sehgal,
Michael Seiffert,
Leonardo Senatore,
Paolo Serra,
Sarah Shandera,
Meir	Shimon,
Peter Shirron,
Jonathan	Sievers,
Kris Sigurdson,
Joe  Silk,
Robert Silverberg,
Eva	Silverstein,
Suzanne	Staggs,
Albert Stebbins,
Federico Stivoli,
Radek Stompor,
Naoshi Sugiyama,
Daniel Swetz,
Andria Tartari,
Max	Tegmark,
Peter Timbie,
Matthieu Tristram,
Gregory Tucker,
Jon	Urrestilla,
John	Vaillancourt,
Marcella Veneziani,
Licia	Verde,
Joaquin Vieira,
Scott	Watson,
Benjamin	Wandelt,
Grant Wilson,
Edward Wollack,
Mark	 Wyman,
Amit	Yadav,
Giraud-Heraud	Yannick,
Olivier Zahn,
Matias Zaldarriaga,
Michael Zemcov,
Jonathan	Zwart

\bigskip
\centerline{February 14, 2009}
\vskip10pt
\parskip=2pt

\newpage

{\bf Executive summary}

\medskip

{\bf How did the universe evolve? The fine angular scale ($\ell>1000$) temperature and 
polarization anisotropies in the CMB are a Rosetta stone for understanding 
the evolution of the 
universe. Through detailed measurements one may address everything from 
the physics of the birth of the universe to the history of star formation 
and the process by which galaxies formed.  One may in addition track the 
evolution of the dark energy and discover the net neutrino mass. }

{\bf We are at the dawn of a new era in which hundreds of square degrees of sky can be mapped 
with arcminute resolution and sensitivities measured in microKelvin. 
Acquiring these data requires the use of special purpose telescopes such 
as the Atacama Cosmology Telescope (ACT), located in Chile, and the South 
Pole Telescope (SPT).  These new telescopes are outfitted with a new 
generation of custom mm-wave kilo-pixel arrays. Additional instruments are 
in the planning stages.}  

\medskip

\section{Introduction}

The primary CMB has been a gold mine for understanding the cosmos. Through the study 
of the CMB we have determined the geometry, age, and contents of the universe at the few percent level. Observations have reached the point where we now have a ``standard model of cosmology."  Yet there is much more to learn from the CMB.  In the standard model there are new unanswered
questions and more traditional questions are more sharply focused. The following are within our reach
in the next decade: 
\begin{enumerate}
\item What is the dark energy and what are its characteristics? Did the  dark energy act differently
before $z=1$?
\item Did neutrinos leave an identifiable imprint on the cosmos and if so what is the sum of their 
masses? Measurements of neutrino oscillations show that the difference in the square of the 
masses is $\sim0.002$~(eV)$^2$, indicating that at least one species must have a mass near $0.05~$eV. This value can be determined with fine scale anisotropy measurements.  
\item Where are the missing baryons? Big Bang nucleosynthesis and CMB derived baryon densities
are not in accord with the observational census. 
\item How did the first stars turn on and what is their ionization history? 
\item Did the early universe have only Gaussian fluctuations or were there phase transitions that perhaps produced cosmic strings? The discovery of primordial non-Gaussianity
would revolutionize cosmology.  
\item Are the fluctuations solely adiabatic or is there an admixture of isocurvature modes? 

\end{enumerate}
All of these questions may be addressed with measurements of the fine angular scale anisotropy in the CMB. The answers to a number of the questions come both from the CMB itself and from the CMB
in correlation with radio, infrared, visible, and X-ray  observations of  galaxies and clusters of galaxies.
Thus there are built in consistency relations. 

Our knowledge to date has been based primarily on the anisotropy at angular scales larger than a quarter degree ($\ell\sim1000$) and with the successful launch of the Planck satellite this will be pushed to 
a tenth degree  ($\ell\sim2500$) .   {\it The fine angular scale measurements
$(1000\ltsim \ell \ltsim10000 )$  described here are a critical complement to Planck.}

At angular scales larger than $\sim 0.1^{\circ}$, the anisotropy may be thought of as a direct probe of the response of the CMB to perturbations laid down in the early universe as seen at a redshift of $z=1090$. The fluctuations are a part in $10^5$ of the background and with linear perturbation theory the properties of the CMB may be computed with exquisite accuracy.

At smaller angular scales, new phenomena become apparent. Objects such as 
galaxies and clusters of galaxies emerge from the primordial plasma and leave their imprint on the CMB. These objects and their environments can in turn be used as beacons with which to interrogate the evolution of spacetime. They can be used to probe components of the standard model such as the 
neutrino mass to new depths. And through a rich set of cross correlations
with radio, infrared, optical, and X-ray surveys, the fine scale anisotropy may be used to pin down the process of cosmic structure formation.

At small angular scales, one may think of the CMB as a backlight with precisely known statistical properties at a precisely known distance. This light illuminates all that is between us and the 
surface of decoupling. Different phenomena leave different imprints on the light.
For example, the hot electrons in galactic clusters 
reveal their presence by scattering the CMB with a characteristic frequency signature. This is called the Sunyaev-Zel'dovich (SZ) effect.  In another mechanism, mass concentrations throughout the universe gravitationally lens the CMB. This lensing can be measured through the correlations it imposes on the CMB. 

A new generation of instruments is poised to increase the angular resolution of CMB-frequency maps
by a factor of five and to increase the sensitivity by an order of magnitude. Because of the vast scientific possibilities,  an even more advanced generation of instruments is already in the design phase. 

\section{The Science}

The fine-scale anisotropy, both temperature and polarization, provide 
four distinct avenues for scientific investigation.  
There has been considerable theoretical work on all the general areas mentioned below and there
are more research topics than there is space to mention.  
Because there are thousands of papers, we limit references to review articles. 
One should also keep in mind that the fine angular scale anisotropy is relatively unexplored.  
With the recent advances in sensitivity, there is significant potential for discovering new phenomena.

\begin{enumerate}
\item {\bf Measure the intrinsic anisotropy to determine the high-$\ell$ tail of the primary anisotropy 
and to search for intrinsic non-Gaussianity.}  Our most direct probe of the infant universe is the scalar spectral index, $n_s$, and its change with scale. The formal accuracy on $n_s$ from the Planck 
satellite is 0.5\%. However, our confidence in  the result will depend on detailed knowledge of the transition from the linear regime (primary CMB) to the non linear regime (secondary CMB). This transition can only be measured through the fine scale anisotropy.  In addition, we will want to be certain that $n_s$ is not being affected by foreground emission, point sources, or low levels of secondary anisotropies. This is best done through fine scale anisotropy measurements. 

In the standard model, the fluctuations are Gaussian. It is widely believed that if the model is incomplete or incorrect the first hints will come through the detection of non-Gaussianity. Alternative models of the early universe and remnants of primordial phase transitions predict measurable levels of non-Gaussianity. For example, cosmic strings would be directly detectable through their imprint on the CMB. Fine scale measurements will provide the crucial results that will build on those from Planck.

\item {\bf Measure the gravitational lensing of the CMB [1].}  The CMB lensing field can be described
by the deflection field as $T_{obs}(\hat{n}) = T_{int}(\hat{n}+\vec d)$ where $T_{int}$ is the 
unlensed temperature field, $\hat{n}$ is the direction, and $\vec d$ is the deflection field.
The deflection field may be reconstructed from the four-point distribution of the temperature 
anisotropy and from the polarization B-modes. {\it Thus the measurement of both the fine scale temperature and polarization anisotropy is important.} The lensing B-modes are distinct from the 
ones associated with inflation and must exist. 

The deflection field is a measure of the effects of the spacetime between us and the decoupling surface and thus probes different physical processes than does the primary anisotropy. 
The deflection field is sensitive to the volume between us and the decoupling surface.
As a consequence, it breaks the ``geometric degeneracy" associated with the primary anisotropy. With the primary anisotropy in hand,
the power spectrum of the deflection field can give an excellent measure of the neutrino mass, spatial curvature, and ``early dark energy."  For example, a measurement of the polarization to a level of $5~\mu {\rm K-arcmin}$ with a $\theta_{FWHM}=2~$arcmin beam over a quarter of the sky has an uncertainty  on the sum of neutrino masses of 0.05 eV and on the curvature of 0.2\%.  CMB lensing 
provides the best way to study the nature of dark energy at early times because the 
lensing kernel probes a wide range of redshifts that peaks at $z \sim 3$ to 4, while low-redshift
cosmological probes, including galaxy lensing, are sensitive to cosmology at $z < 2$.
These measurements are well within our reach in the coming decade.

The deflection field may also be correlated with the SZ effect, galaxy shear, the LRGs and a host of
other phenomena to find the growth rate of structure. The growth rate in turn is another probe of
dark energy and the mass of the neutrino [2].

\item {\bf Find clusters of galaxies through their SZ effect, determine the cluster redshifts with 
optical follow up, understand the mass selection function with a combination of
SZ, optical, and X-ray measurements, and from the cluster catalog
determine $dN/dz$ or $dN(>M)/dM$} [3]. The number distribution is exponentially sensitive to 
the dark matter and dark energy densities. Depending on the visibility of clusters,
the equation of state may be determined to $\sim10$\% accuracy. 
Multi-frequency observations  to separate the thermal SZ from the kinetic SZ and primordial fluctuations
will be an important component of this research.
This program will evolve throughout the decade and is anticipated to be an important complement to other methods based on weak lensing, supernovae, and baryon acoustic oscillations.

There are a host of other phenomena one may pursue. A key attribute of a survey of SZ clusters 
is a well defined selection function that is almost redshift independent. The sample may 
be used to constrain the neutrino mass and is especially sensitive to $\sigma_8$.
Ultimately, it may even be possible to measure large scale structure through the cluster polarization.  

\item  {\bf Correlate and compare the CMB with lower redshift cosmological measurements.}
In the most straightforward application,  one uses the interaction of the CMB with lower redshift phenomena and from that determines the growth rate of structure. The growth rate is then directly related to cosmology.  However, it is also possible to 
examine other phenomena.  For example, many believe that the ``missing baryons" 
are in the outskirts of clusters. If this is the case, they should be visible by stacking clusters
and identifying the SZ effects. In another example,  the correlations are a probe of the process of reionization. And in yet another, the cross-correlation with massive galaxies 
has the potential to measure the energy feedback from supermassive black holes, which heats the surrounding intergalactic medium and creates a small-scale SZ distortion. 
The various phenomena are distinguishable through their specific correlations and spatial distributions.

\end{enumerate}
 
\section{Scientific Instrumentation}
The quest to understand the cosmos through measurements of the CMB has led to
rapid advances in scientific instrumentation and experimental technique. A decade ago, researchers
were talking about arrays of hundreds of detectors and observing with a handful. Today observations are made with kilo-pixel arrays, and larger and more sophisticated arrays are in the works.
The field has been an incubator for multiple new and diverse technologies.

In the past two years two telescopes dedicated to measuring the fine scale anisotropy have
been commissioned and have begun taking data. One is the Atacama Cosmology Telescope
(ACT) located in northern Chile near ALMA and the other is the South Pole Telescope (SPT) 
located at the Amundsen-Scott South Pole station. The telescopes complement each other in their sky coverage, technology, and observing techniques. They have also observed a common area of sky
for cross calibration.

Table 1 gives a list of instruments that are anticipated to be measuring the fine
angular scale anisotropy and the SZ effect in the next decade. It is not exhaustive.
It omits some efforts early in the proposal phase. For example, a balloon borne
arcminute CMB measurement has been discussed. It also omits a good number of CMB 
experiments targeted at measuring the polarization in the $\ell<2000$ range. These 
experiments are discussed in a separate report authored by Stephan Meyer.  

\begin{table}[h]
\caption{Telescopes for measuring the $\ell>1000$ CMB anisotropy.}
\begin{center}
\begin{tabular}{lrccc} 
\hline Name & Location & Diameter/Separation & Wavelength & $\ell_{max}$ \\
& & (m) & (mm) &  \\
\hline 
ACT &	Atacama  & 6	&1-2	& 7800$^{1}$\\
PolarBear	& Atacama  & 3.5	& 1-3	& 3000$^{1}$ \\
SPT &	South Pole	& 10 	& 0.3 - 3	& 9000$^{1} $\\
\hline
AMI$^2$ &	MRAO, UK & 3.7-13	&18	&  10000\\
AMiBA $^2$ &	Hawaii & 1	& 3	&  $>$2000\\
APEX-SZ$^3$ &	Atacama & 12	&1-3	&  12000$^{1}$\\
AzTEX on the LMT  &	Sierra Negra, MX & 50 & 0.7-3&  60000\\
CCAT$^4$ &	Atacama & 25	&0.7-3&  40000\\
MUSTANG on the GBT$^3$ & West Virginia & 100 & 3&  80000\\
SZA/CARMA$^5$ & California	& 3.5-10.4  & 3-10 & $>$10000\\
\hline 
\multispan{5}  \leftline{ The lower entries are anticipated to focus on SZ and source 
measurements in the 2010 decade.} \\    
\multispan{5}  \leftline{ One should also keep in mind that observation wavelength is flexible.} \\    
\multispan{5}  \leftline{ $^1$ Taken as $\pi/\theta_{\rm FWHM}$ at 150~GHz.} \\    
\multispan{5}  \leftline{ $^2$ Interferometer with resolution dependent on baseline.} \\    
\multispan{5}  \leftline{ $^3$ General purpose. Observes the CMB/SZ only part of the time.} \\    
\multispan{5}  \leftline{ $^4$ Proposed for SZ cluster and other studies.} \\
\multispan{5}  \leftline{ $^5$ Heterogenous interferometer array for SZ imaging.} \\
\hline \end{tabular} \end{center} \label{tab:telescopes} 
\end{table}

\begin{figure}[h]
\begin{minipage}[b]{7.0cm}
\begin{center}
\includegraphics[width=7.0cm]{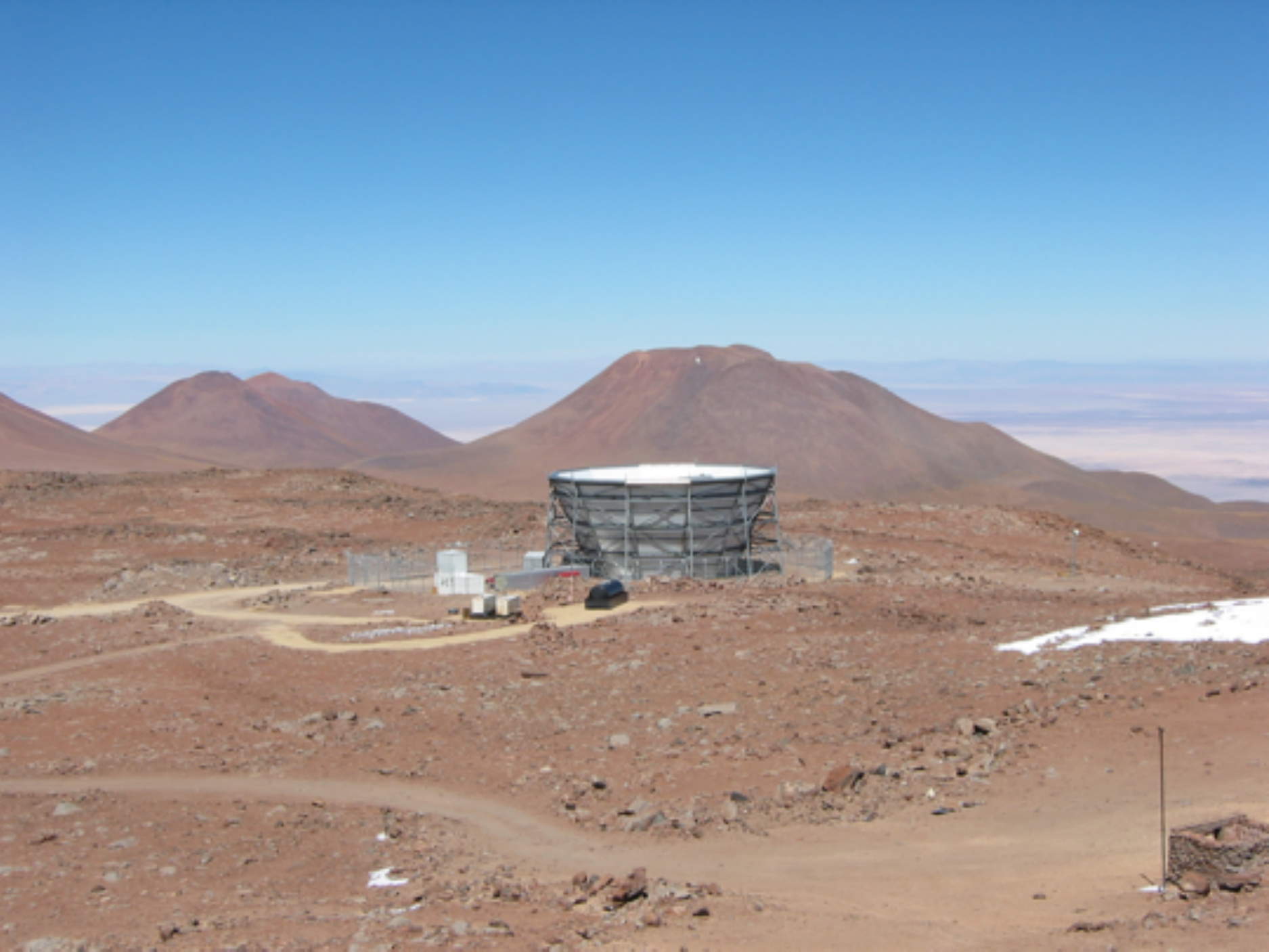}
\end{center}
\end{minipage}
\hfill
\begin{minipage}[b]{8.5cm}
\begin{center}\
\includegraphics[width=8.25cm]{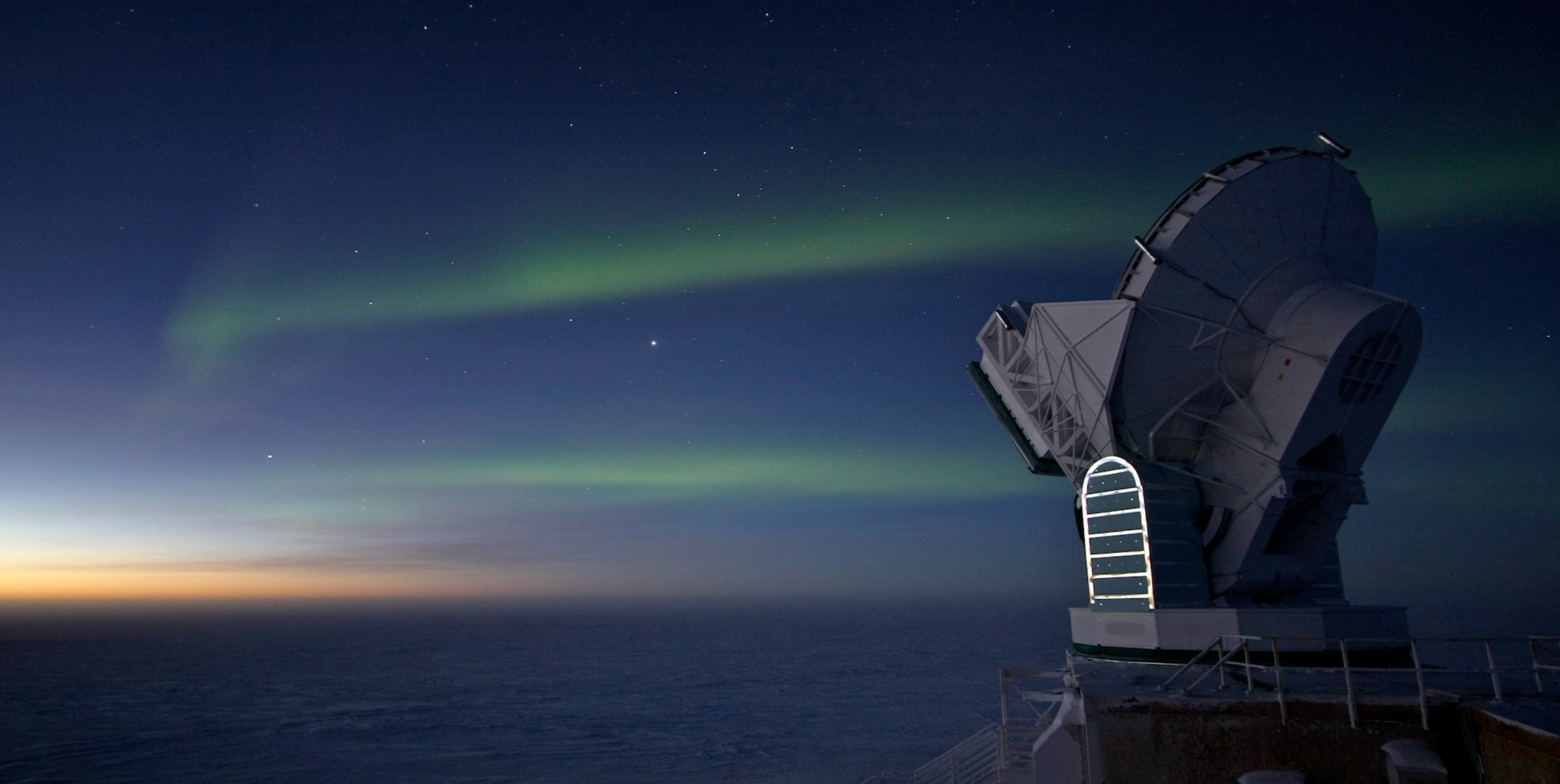}
\end{center}
\end{minipage}
\caption[]{\label{fig:telescopes}\small On the left is the Atacama Cosmology Telescope
which is situated in the Chajnantor Science Preserve in northern Chile, adjacent to the ALMA site.
The entire telescope is inside a three story high ground screen. ACT is at 5200~m. It takes 
advantage of the high dry site. On the right is the South Pole Telescope with the aurora in the background. SPT is at 2800~m, with an equivalent pressure altitude of 3500-4000~m. It takes advantage of the flat and exceptionally dry  polar cap.  Photo credits Adam Hincks (ACT) and  Keith Vanderlinde (SPT). }
\end{figure}

\begin{figure}[h]
\begin{minipage}[b]{7.5cm}
\begin{center}
\includegraphics[width=7.5cm]{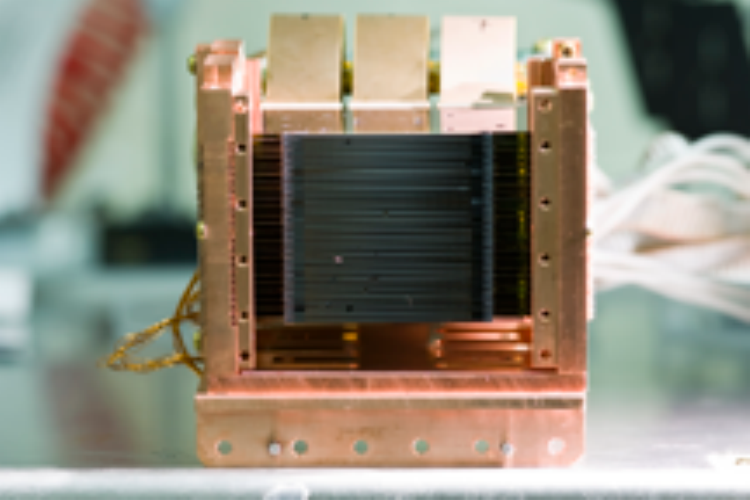}
\end{center}
\end{minipage}
\hfill
\begin{minipage}[b]{7.0cm}
\begin{center}\
\includegraphics[width=7.0cm]{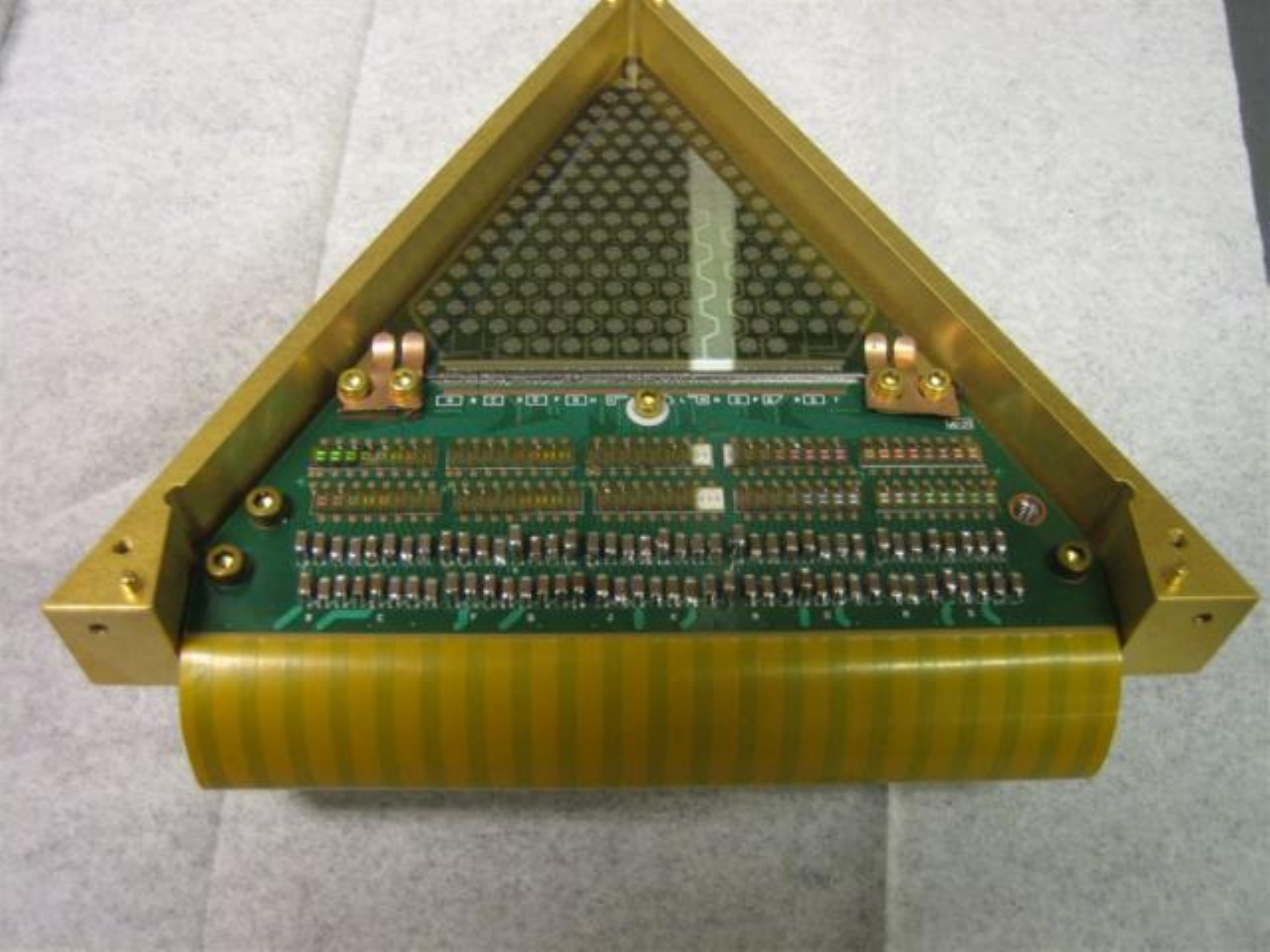}
\end{center}
\end{minipage}
\caption[]{\label{fig:dets}\small Detector arrays for ACT and SPT. On the left is one of the ACT arrays. It contains 32 by 32 detectors that were fabricated at NASA/GSFC. Cryogenic optics images the sky onto the 32 by 33 mm active area. Each strip of 32 detectors is built onto a silicon card. 
On the right is an image of a  SPT sub array developed at Berkeley. The TES detectors are at the top.
Feed horns (not shown) couple the radiation onto the detectors.
The circuit board at the bottom contains multiplexing circuitry. }
\end{figure}

\begin{figure}[t]
\begin{minipage}[b]{7.5cm}
\begin{center}
\includegraphics[width=7.5cm]{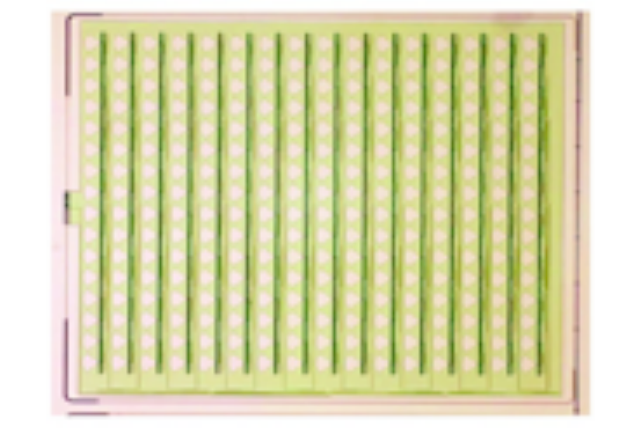}
\end{center}
\end{minipage}
\hfill
\begin{minipage}[b]{7.0cm}
\begin{center}\
\includegraphics[width=7.0cm]{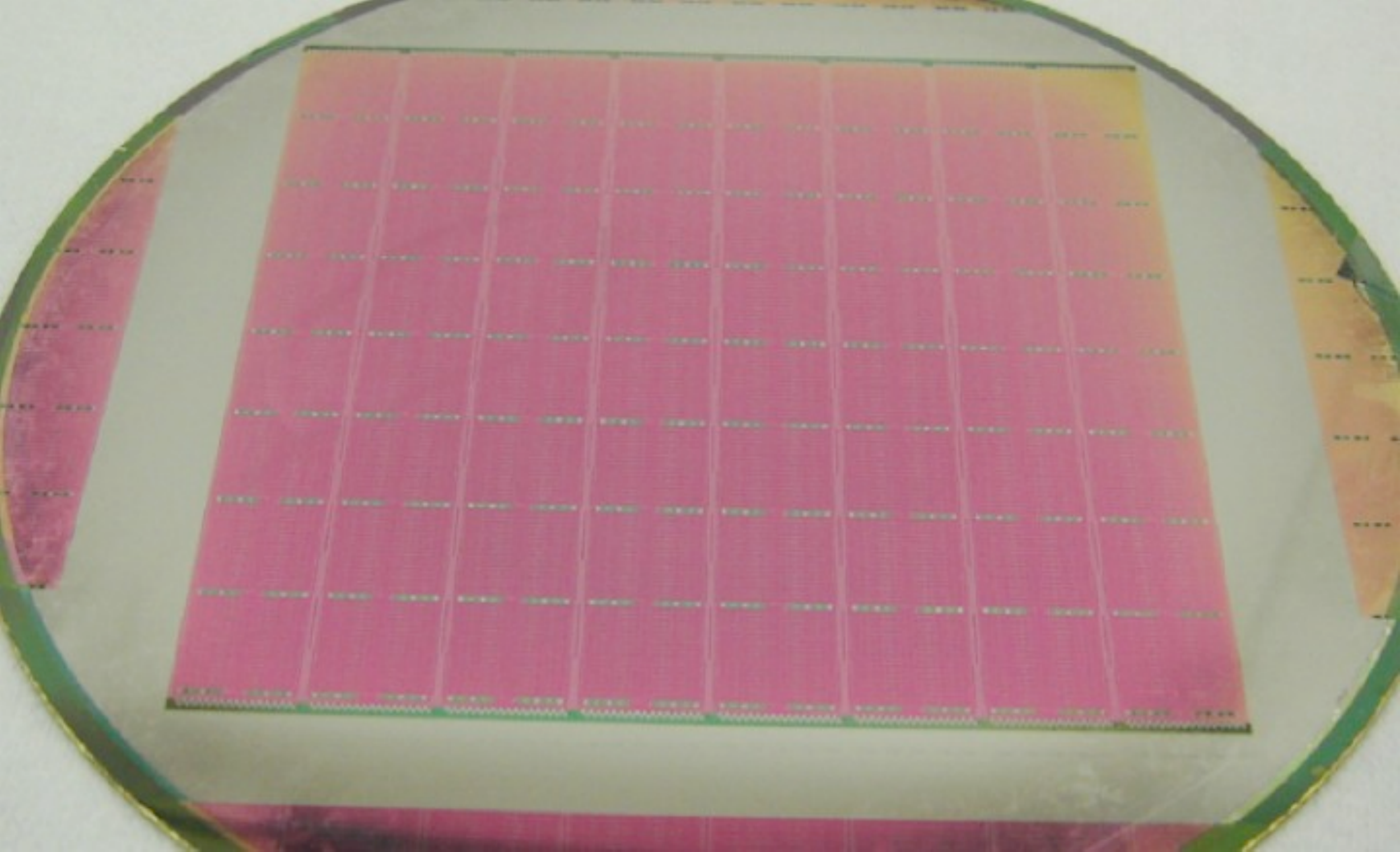}
\end{center}
\end{minipage}
\caption[]{\label{fig:dets2}\small On the left is a MKID demonstration array. On the right is a planar phased array of antennas coupled to detectors. Each of the 64 elements measures both polarizations. Multiple sub arrays may be combined. Both of these examples are from Caltech/JPL.}
\end{figure}

New detector arrays have been developed to operate with the new telescopes. Superconducting
transition edge bolometers operating near 0.3 K are currently the detectors of choice.
Figure~\ref{fig:dets} shows two examples. Other arrays based
on microwave kinetic induction detectors, MKIDs, and planar phased array antennas are also under development. Examples are shown in Figure~\ref{fig:dets2}. The large number of detectors has required new readout and multiplexing electronics. Both time domain multiplexing, developed at NIST, and frequency domain multiplexing, developed at Berkeley, are being employed.  

\section{References}
 
\noindent [1]``CMBPol Mission Concept Study: Gravitational  Lensing,"
K. Smith et al.,  arXive:0811.3916v1,  Nov 2008.

\noindent [2]``Findings of the Joint Dark Energy Mission Figure of Merit Science Working Group,"
Albrecht et al. 2009, eprint arXiv:0901.0721.

\noindent [3]``Cosmology with the Sunyaev-Zel'dovich Effect," Carlstrom, J., Holder, G., \&  Reese, E.,
Annual Review of Astronomy and Astrophysics, Vol. 40, p. 643-680, 2002.

\vfill

\end{document}